\documentclass[amsmath,amssymb,twocolumn]{revtex4-2}

\usepackage{graphicx}
\usepackage{cleveref}
\begin{document}

\title{Reply to the Comment by Hradil and Řeháček}

\author{Jens A. H. Nielsen}
\author{Jonas S. Neergaard-Nielsen}
\author{Tobias Gehring}
\author{Ulrik L. Andersen}
\affiliation{Center for Macroscopic Quantum States (bigQ), Department of Physics, Technical University of Denmark}

\begin{abstract}
    Hradil and Řeháček argue that the quantum metrological advantage demonstrated in our squeezed-vacuum phase-sensing experiment disappears, when measurement repetitions and total photon resources are properly accounted for. We show that this conclusion is mathematically unfounded and that neither resource accounting nor finite-sample statistical considerations invalidate the demonstrated quantum-enhanced local phase sensitivity.
\end{abstract}

\maketitle

In two recent manuscripts~\cite{Hradil2,Hradil1}, Hradil and Řeháček challenge the interpretation of the quantum metrological advantage demonstrated in our experiment on squeezed-vacuum phase sensing~\cite{Nielsen}. Here, we address their criticisms and identify several incorrect conclusions and mischaracterizations of our experimental analysis.

The first manuscript titled "Fundamental and operational limitations of Fisher-information-based quantum metrology" \cite{Hradil2}, discusses the general limitations of using quantum Fisher information as a benchmark for metrological performance. In particular, the authors argue that large Fisher information does not necessarily translate into an operational quantum advantage when finite-sample statistics and total resources are considered. They apply this reasoning to our squeezed-vacuum phase-estimation protocol, using the chi-squared statistics of the measured quadrature variance to question the claimed enhancement.

The second manuscript \cite{Hradil1}, a Comment on our work, applies these arguments more directly to our experiment. Its central claim is that the quantum advantage disappears when the number of measurement repetitions is included in the total photon-resource budget. The authors further question the operational attainability of the predicted sensitivity and the role of prior phase information. Below, we address these three issues and demonstrate that, although some of the general statistical considerations are valid, the conclusions drawn about our experiment are not. 

{\it First, the Comment incorrectly associates the normalization of Fisher information per probe photon with the omission of measurement repetitions.} For a pure single-mode squeezed vacuum with mean photon number $\langle n\rangle=\sinh^2r$, the maximum classical Fisher information (FI) attainable by homodyne detection is $F_{\mathrm{sqz}}=8\langle n\rangle(\langle n\rangle+1)$, which equals the quantum Fisher information (QFI) in the ideal lossless case. For $M$ statistically independent measurements, the total FI is $F_{\mathrm{tot}}=MF_{\mathrm{sqz}}$, while the total number of probe photons is $N_{\mathrm{tot}}=M\langle n\rangle$, precisely as required by Eqs. (1)-(3) of the Comment. However, an optimally measured coherent-state probe with the same photon number per measurement yields $F_{\mathrm{coh,tot}}=4M\langle n\rangle$. Consequently,
\begin{equation}
\frac{F_{\mathrm{sqz,tot}}}{F_{\mathrm{coh,tot}}}
=\frac{8M\langle n\rangle(\langle n\rangle+1)}{4M\langle n\rangle}
=2(\langle n\rangle+1)>1.
\end{equation}
Thus, including all measurement repetitions does not remove the quantum enhancement: the factor $M$ cancels when comparing strategies using equivalent photon resources and local estimation tasks. Equivalently, the relevant FI per total probe photon is $F_{\mathrm{tot}}/N_{\mathrm{tot}}=F/\langle n\rangle$, independent of $M$. Optical losses reduce the achievable FI, as explicitly considered in Ref.~\cite{Nielsen}, but do not alter this resource-accounting principle. The inference in the Comment that accounting for repeated measurements necessarily restores classical performance is therefore incorrect.

\begin{figure*}
    \centering
    \includegraphics[width = 1. \linewidth]{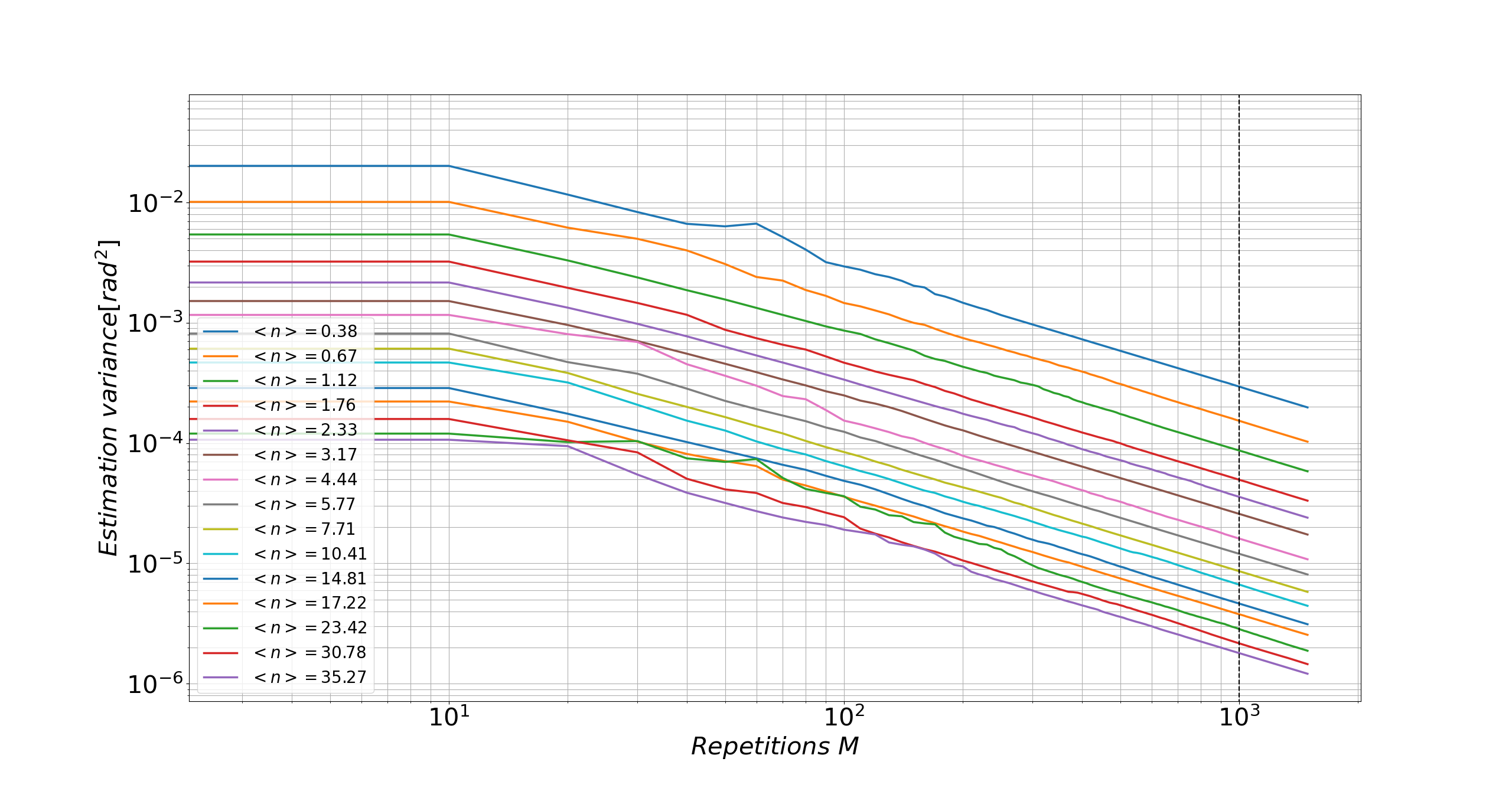}
    \caption{Convergence of the Bayesian estimation variance with the number of homodyne measurements (M), for the different mean photon numbers investigated in Ref.~\cite{Nielsen}. Each curve shows the estimation variance experimentally obtained from the first (M) measurements at the corresponding operating point. The vertical dashed line indicates (M=1000), the number of measurements used for all phase-estimation results in Fig. 3 of Ref.~\cite{Nielsen}.}
    \label{fig:conv}
\end{figure*}

{\it Second, our experiment did not infer phase precision solely from theoretical QFI.} As explicitly described in Ref.~\cite{Nielsen} and repeated in \cite{Hradil1}, we implemented Bayesian phase estimation using $M=1000$ homodyne measurements for each static phase setting and $M=100$ measurements for dynamic phase tracking. The likelihood of the complete dataset,
\begin{equation}
p(\{x_i\}_{i=1}^M\mid\phi)=\prod_{i=1}^{M}p(x_i\mid\phi),    
\end{equation}
was used to determine the posterior phase distribution, its maximum, and its associated uncertainty. The classical Fisher Information was extracted from the estimation uncertainty. The resulting phase estimates and their convergence with increasing measurement number were experimentally investigated, as illustrated in Figs. 2-4 of Ref.~\cite{Nielsen}. Consequently, the criticism that our conclusions rely exclusively on single-measurement QFI while neglecting the measurements required to construct an estimator mischaracterizes our experimental analysis. Although finite-sample estimator bias and mean-squared error are distinct from posterior uncertainty, these considerations do not establish that the observed local enhancement disappears. To further examine the finite-sample behavior, we have reanalyzed the estimation variance as function of the number of measurements, $M$, for all investigated squeezing levels, as shown in \cref{fig:conv}. The variances converge and stabilize to a $1/M$ scaling well before the $M=1000$ point and maintain the same scaling even at higher squeezing levels. Importantly, all results in Fig. 3 of Ref.~\cite{Nielsen} were obtained using the same fixed number of measurements, $M = 1000$, independently of the squeezing level. Thus, the experimentally observed improvement in phase sensitivity with increasing squeezing was not obtained by increasing the number of measurement repetitions, contrary to the mechanism suggested in the Comment.Furthermore, each point in figure 3 (c) of Ref.~\cite{Nielsen} represents an average over 100 independent estimation runs, each comprising $M = 1000$ measurements, with statistical error bars smaller than the plotted markers.

{\it Third, the chi-squared statistical argument presented in Ref.~\cite{Hradil2} does not invalidate the squeezing-dependent phase sensitivity.}
For independent Gaussian homodyne outcomes with phase-dependent variance $V(\phi)$, the statistic $Y=\sum_{i=1}^{M}x_i^2$ satisfies $Y/V(\phi)\sim\chi_M^2$. Although the relative variance-estimation uncertainty scales as $M^{-1/2}$, the corresponding phase sensitivity depends also on the phase response of the variance. Indeed, the exact Fisher information contained in $Y$ is
\begin{equation}
F_Y(\phi)=\frac{M}{2}
\left[\partial_\phi\ln V(\phi)\right]^2
=MF_{\mathrm{hom}}(\phi).
\end{equation}
Thus, the chi-squared statistic retains the full squeezing-dependent phase Fisher information for any $M$. The statistical scaling of the variance estimator with measurement number does not imply classical scaling of the phase sensitivity with probe photon number. While finite-sample estimator performance requires separate verification, the chi-squared argument does not establish the disappearance of the quantum advantage.

{\it Finally, the distinction between local and global phase estimation is important but does not contradict our experimental conclusions.} Our protocol is designed for high-precision measurement of small phase variations around an optimal operating point, rather than unrestricted estimation of an arbitrary unknown phase. This limitation was explicitly acknowledged in Ref.~\cite{Nielsen}, and phase locking was employed in our dynamic sensing demonstration. Deterministic operation refers to the absence of probabilistic state preparation and measurement post selection, not to the absence of a phase reference or prior information. Likewise, our comparisons with ideal N00N states and the conventional Heisenberg limit concern local phase sensitivity under the photon-resource convention specified in Ref.~\cite{Nielsen}, rather than universally attainable global estimation precision. Phase initialization and locking may require additional resources in a fully operational comparison, but equivalent prior information and resource conventions must be applied to classical and quantum protocols.

In conclusion, neither consistent accounting of repeated measurements nor the chi-squared statistics of finite homodyne datasets establishes the disappearance of the quantum-enhanced local phase sensitivity demonstrated in Ref.~\cite{Nielsen}.

\end{document}